\newif\ifSC

\SCtrue
\ifSC
\documentclass[onecolumn,draftclsnofoot,12pt]{IEEEtran}
\else
\documentclass[10pt,twocolumn]{IEEEtran}
\fi
\usepackage{cite}
\usepackage{amsmath,amssymb,amsfonts,amsthm}

\usepackage{algorithmic}
\usepackage{graphicx}
\usepackage{textcomp}
\usepackage{color}
\usepackage{varioref}
\usepackage{etoolbox}
\usepackage{xcolor}
\usepackage{bbm}
\usepackage{lipsum}
\usepackage[normalem]{ulem}
\usepackage{mathtools}
\usepackage{subfigure}
\def\home{\hbox{\kern3pt \vbox to13pt{}%
		\pdfliteral{q 0 0 m 0 5 l 5 10 l 10 5 l 10 0 l 7 0 l 7 5 l 3 5 l 3 0 l f
			1 j 1 J -2 5 m 5 12 l 12 5 l S Q }%
		\kern 13pt}}

\newtheorem{theorem}{Theorem}

\newtheorem{corollary}{Corollary}
\newtheorem{remark}{Remark}
  
  \usepackage{mathtools} 
  
\def\BibTeX{{\rm B\kern-.05em{\sc i\kern-.025em b}\kern-.08em
		T\kern-.1667em\lower.7ex\hbox{E}\kern-.125emX}}
\newcounter{relctr} 
\everydisplay\expandafter{\the\everydisplay\setcounter{relctr}{0}} 

\newcommand\numeq[1]%
{\stackrel{\scriptscriptstyle(\mkern-1.5mu#1\mkern-1.5mu)}{=}}
\AtBeginDocument{}

\newcommand{\x}{\mathbf{x}}
\newcommand{\X}{\mathbf{X}}
\newcommand{\Y}{\mathbf{Y}}
\newcommand{\y}{\mathbf{y}}

\newcommand{\St}{\mathsf{S}}
\newcommand{\sen}{\mathrm{S}}
\newcommand{\1}{\mathbbm{1}}

\newcommand{\drm}{\mathrm{d}}
\newcommand{\uni}{\mathrm{u}}

\newcommand{\f}{\mathrm{f}}
\newcommand{\ob}{\mathrm{o}}

\newcommand{\F}{\mathrm{F}}

\newcommand{\Nt}{\mathrm{N}}
\newcommand{\Ct}{\mathrm{C}}
\newcommand{\dv}{\mathrm{d}}

\newcommand{\B}{\mathcal{B}}

\newcommand{\st}{\alpha(r)}

\newcommand{\A}[3]{\mathcal{A}(#1,#2,#3)}
\newcommand{\C}{\mathcal{C}}

\DeclareMathOperator\erf{erf}
\newcommand{\ths}{\mathrm{th}}
\newcommand{\ie}{{\em i.e.}}
\newcommand{\dist}[1]{\|#1\|}
\newcommand{\indside}[1]{\mathbbm{1}\left(#1\right)}
\newcommand{\insertnotationtable}{
	
	\begin{table}
		\caption {\small Notation Table} \label{tab:title}
		\begin{center}
			\begin{tabular}{|p{.9in} | p{2.23in} |}
				\hline 
					\textbf{Symbol} & \textbf{Definition}  \\ \hline 
					$\B(\X,r)$ & Ball of radius $r$ centred at $\X$.\\ \hline
				$\Phi_\f$ & 2D FHPPP modeling location of sensors in the forest \ie $\B(\ob,r_\drm)$ \\ \hline
				$\lambda_\f$ & Density of wireless sensor network (number of sensors per unit area).\\ \hline
				$r_\sen$& The fixed sensing radius of each sensor.\\ \hline
				$\X_i$ & The location of the $i$th sensor.\\\hline 
			
				$\mathsf{S}_i$ & Sensing region of $i\ths$ sensor \ie $ \mathcal{B}(\ob,r_{\sen})$\\\hline
				$\xi$ & Total sensing area of all sensors.\\ \hline
				$\A{r}{r_\drm}{z}$ & Area of intersection between two circles located $z$ distance apart with  radii $r$ and $r_\drm$.\\ \hline
				$\oplus$ & Minkowski addition operator.\\ \hline
				$Y=\|\Y\|$ & L-2 norm of $\Y$.\\ \hline
				$\Phi_{\f}(\mathsf{A})$ & Random variable denoting the total number of points of $\Phi_{\f}$ located inside  the set $\mathsf{A}$.\\ \hline
				$\B(\ob,r_\drm)\cap \B(\Y,r)$ & Intersection between the circles  $\B(\ob,r_\drm)$ and $\B(\Y,r)$.\\ \hline
						$|\mathsf{A}|$ & Area of any set $\mathsf{A}$.\\ \hline
			\end{tabular}
		\end{center}
		
\end{table}}

\begin{document}

\title{On Detection of Critical Events in a Finite Forest using Randomly Deployed Wireless Sensors}

\author{\IEEEauthorblockN{ Kaushlendra K. Pandey, Abhishek K. Gupta}\\
\IEEEauthorblockA{\textit{Department of Electrical Engineering} \\
\textit{Indian Institute of Technology Kanpur, India} \\
Email: kpandey@iitk.ac.in, gkrabhi@iitk.ac.in}
}

\maketitle

\begin{abstract}
Ecosystem of a forest suffers from many adverse events such as wild-fire which can occur randomly anywhere in the forest and grows in size with time.
This paper aims to analyze performance of a network of randomly deployed wireless sensors for the early detection of these time-critical and time-evolving events in a forest. We consider that the forest lies in a confined space ({\em e.g.} a circular region) and the wireless sensors, with fixed sensing range, are deployed within the boundary of forest itself.  
The  sensing area of the network  is modeled as a finite Boolean-Poisson model. In this model, the locations of sensors are  modeled as a finite homogeneous Poisson Point Process (PPP) and the sensing area of each sensor  is assumed to be a finite set. 
This paper aims to answer  questions about the proximity of a typical sensor from a randomly occurred event and the total sensing area covered by  sensors. We first derive the distribution of  contact distance of a FHPPP and the expression of the capacity functional of a finite Boolean-Poisson model. Using these, we then derive the probability of sensing the event at time $t$, termed event-sensing probability. 

\end{abstract}

\section{Introduction}
The ecosystem and biodiversity of forests suffer from various natural and artificial events including wild-fires and spread of disease  \cite{PanGup2018,molina2019analysis}. 
The past literature has suggested that such events can be controlled to avoid severe loss by designing a mechanism for early detection of these  events.
One way to build an efficient alarm system for early detection of these time-critical events is by using wireless sensor network (WSN) of fire sensors deployed in forest. Wireless sensor networks refers to network of sensors connected via wireless links. WSNs are regarded as a cost-effective and inexpensive solution to jointly detect an event/events including fire over an area. However, while deploying a sensor network, it is important to understand the impact of the sensor density on the proximity  of these critical events and the sensing region covered by these sensors in order to ensure that events are detected well in time.

The coverage performance of a WSN with sensors having fixed disk sensing range is analyzed in \cite{kasbekar2011lifetime}. Readers are advised to refer to  \cite{simplot2005energy} for an extensive literature survey discussing the coverage and connectivity analysis of WSNs.
The performance of a WSN can be characterized via various metrics, {\em e.g.} the distance of the closest sensor from an event (termed contact distance) or the probability that the event is sensed by at least one node of the WSN (termed event sensing probability).
The deterministic deployment of sensor nodes may not be possible in the forests where  terrains are not uniform. In these scenarios, sensors are generally deployed randomly. Tools from stochastic geometry provide a tractable framework to study the coverage of random networks including WSN \cite{AndGupDhi16}.
The coverage performance of infinite random WSNs was well studied in the past literature \cite{BaccelliBook}. Modeling of wireless sensor nodes as point process can be justified owing to their random deployment and connectivity mechanism \cite{haenggibook}.   Poisson point process (PPP) is one example of point process which has been widely used in the past literature owing to their tractability. For example, coverage analysis
of wireless sensor network modeled as PPP is performed in \cite{wan2006coverage}. In \cite{kwon2013random}, PPP is used to model a sensor network with sensors acting as data collectors and transmitters to evaluate performance of this network.  The coverage analysis of WSN with deterministic sensing range of individual sensors was performed in \cite{liu2013dynamic} by modeling this network using Boolean-Poisson model \cite{chiu2013stochastic}. Although sensors are moving with time, authors have only considered a particular time snapshot where sensors locations form a PPP. 
Other point processes such as binomial point process (BPP), finite PPP are also used in the past literature. 
For example, WSN has been modeled as BPP.  In \cite{haenggi2007modeling}, the authors presented a closed-form analytical expression for the moment generating function of the interference at the origin. In \cite{srinivasa2010distance}, authors presented the closed form expression for the different distance distribution of BPP. In \cite{bettstetter2003hop}, authors studied the distance distribution in a multi-hop network with $n$ nodes uniformly distributed in a square.

The expression for CDF of contact and nearest neighbor distance for homogeneous infinite PPP is available in \cite{AndGupDhi16}. The expression for capacity functional for a homogeneous PPP is derived in \cite{chiu2013stochastic}. 
Since the forest lies in finite space, the deployed sensor network is also finite. There has been past research to investigate  distribution of various distances among nodes that are uniformly distributed in a confined space.  
In \cite{miller2001distribution}, the authors derived the cumulative distribution function (CDF) of the distance between two randomly located mobile devices. In \cite{tseng2006distance}, authors have considered wireless nodes to be uniformly located within a square and calculated the  distribution of the distance of the $k$-th neighbor along with the distance  between two randomly selected nodes. 
 The performance of a sensor network modeled as a Boolean process with nodes located as PPP in a confined space to detect a dynamic and time-evolving event, for example, its coverage probability and the closest distance of the nearest sensor of this network from an arbitrary point in the same space was not studied in the past work which is the main focus of this paper.

 \begin{figure}[ht!]
 	\def\svgwidth{\textwidth}
 	\centering
 	\includegraphics[width=.48\textwidth]{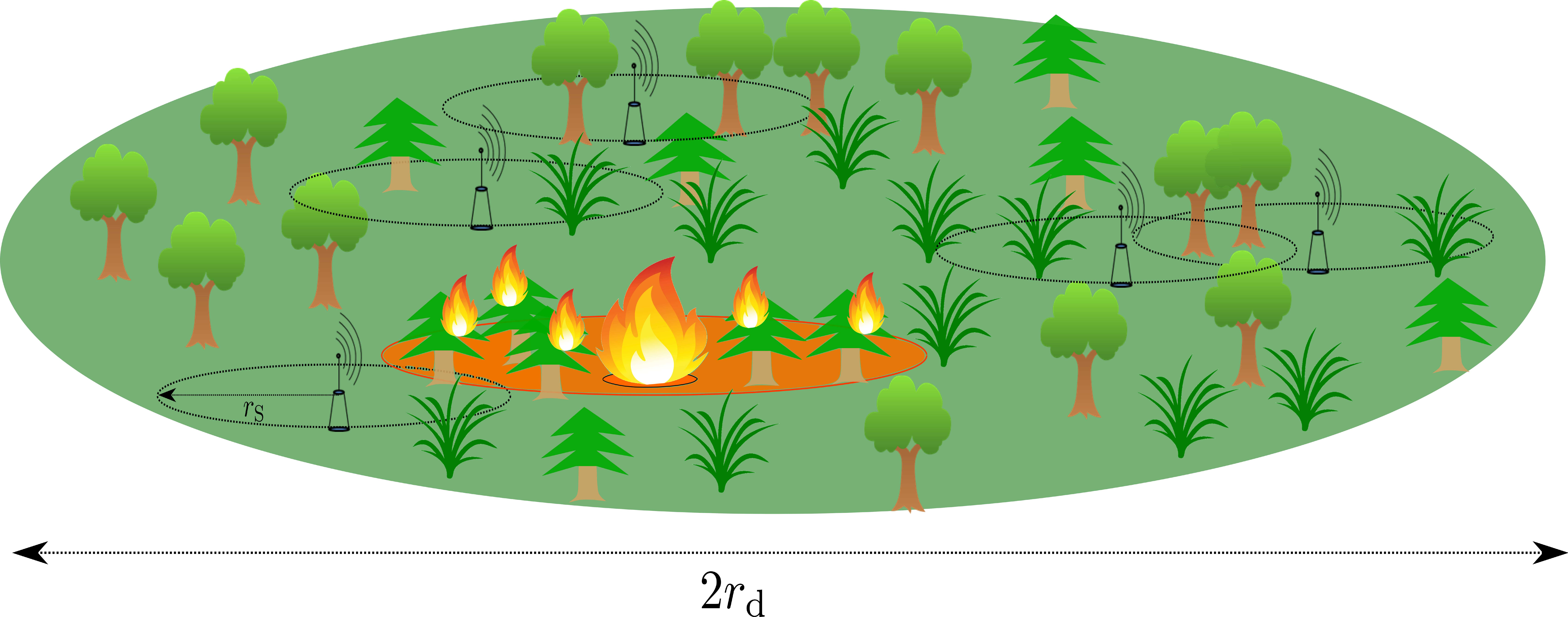}
 	\caption{Illustration of a wireless sensor network deployed in a forest. The forest is located inside a circular area of radius $r_\drm$. Sensors are deployed randomly in the forest. An event of fire has occurred at a location $\mathbb{Y}_\mathrm{u}$. 
 	The outer envelop of fire have adopted a circular shape with a radius increasing with time.
 	} \label{junglefire}
 \end{figure}
 In this paper, we consider a finite wireless sensor network with sensors  in a confined circular forest. 
 Each sensor node has an associated sensing range. The completed sensing range of the network  is modeled as a finite Boolean-Poisson model. In this model, the locations of sensors (termed germs) are  modeled as a finite homogeneous PPP (FHPPP) and the sensing area of each sensor (termed grain) is assumed to be a finite set. 
 We first assume that the target critical event to be sensed/detected can occur uniformly anywhere in the forest. We then compute the distribution of distance of the nearest sensor from the occurrence point of this event. We also derive the expression for the capacity functional of finite Boolean model which help us derive the probability that the event is sensed by at least one sensor at time $t$.

\section{System model}

This paper presents the coverage-analysis of time-critical and dynamic events ({\em e.g.} wild-fires) occurred anywhere in the forest with the help of wireless sensors deployed randomly in the forest. The list of important symbols and notation is represented in table \ref{tab:title}. 

We assume that the forest is  modeled as a 2D ball $\B(\ob,r_\drm)$ with center at the origin and radius $r_\drm$.  The sensor nodes have a fixed sensing radius $r_{\sen}$ around them and are deployed to sense any time-critical event which can occur randomly and uniformly in the forest. We model the coverage area of the sensor network as a Boolean-Poisson model $\xi$. In this process, the locations of sensor are modeled as a FHPPP $\Phi_{\f}=\{\X_i\}$  with intensity \[\lambda(\x)=\frac{m}{\pi r_\drm^2}\1(\dist{\x}\leq r_\drm),\]
 where $m=\lambda(\ob)\pi r_\drm^2$ is the mean number of sensors deployed in the forest. Here, $\X_i$ denotes the location of $i\ths$ sensor. 
Let $\lambda_\f=\lambda(\ob)$ denote the density of this PPP inside the ball $\B(\ob,r_\drm)$. 
Although this paper considers the forest to be confined in a circular region but the result obtained in the paper can be extended to analyze performance in forests of any arbitrary  shape. Since, we have assumed that each sensor have a fixed sensing range $\mathsf{S}_i$,  the occupied space by the sensor network is given by:
\begin{align}
\xi=\bigcup_{\X_i\in\Phi_{\f}} \X_i + \mathsf{S}_{i}.
\end{align}
\textit{Modelling events and their time-evolution}: Let the occurrence/starting/center point of a target event be denoted by $\Y_{\uni}$. The location of $\Y_{\uni}$ is assumed to be uniformly distributed in the forest $\B(\ob,r_\drm)$. 
In other words, the probability density function of the distance of $\Y_{\uni}$ from the origin is given by
\begin{align*}
f_{\dist{\Y_{\uni}}}(y)=\frac{2y}{r_\drm^2}\1({y \le r_\drm}).
\end{align*} 
As stated earlier, we consider events that are time-evolving (in particular growing with time). At time $t$, the event's envelop is denoted by set $K_{\Y_\uni}(t)$ that contains the event occurrence point $\Y_\uni$. The envelop size and shape at time $t$ will depend on the event type and its propagation/evolution characteristics. Consider an example of wild-fires in the absence of wind. Once a fire has occurred at the  location $\Y_{\uni}$, fire will increase in all directions with constant speed $v_{\F}$. Hence, at time $t$, the fire envelop will be a circle, expressed as $\B(\Y_{\uni},v_{\F} t)$.
 \insertnotationtable

\section{Distance Distributions}
In this section, we will derive the CDFs of contact distance of the event and the nearest neighboring sensor distance of a typical point for the FHPP $\Phi_{\f}$.

\subsection{CDF of the contact distance of the event}
The contact distance $R_{\Ct}$ of the event  is a random variable denoting the distance of closest sensor $\{\X_i :\X_i\in\Phi_{\f}\}$ from the event occurrence point $\Y_{\uni}$. It has been assumed that $\Y_{\uni}$ is independent of $\Phi_{\f}$. 
In other words, $R_{\Ct}$ is the distance of the nearest point of a finite homogeneous PPP from a reference location uniformly located inside the range of the finite PPP. 

Let us first condition on the location of $\Y_{\uni}$. The conditional CDF  of the event contact distance $R_{\Ct}(\Y_\uni)$ at $r$ is the probability that nearest sensor from the location $\Y_{\uni}$ is at distance less than or equal to $r$. Mathematically,
\begin{align*}
F_{R_{\Ct}|\Y_{\uni}}(r)=\mathbb{P}(R_{\Ct}(\Y_\uni)\leq r|\Y_{\uni}).
\end{align*} 
Let the notation $\Phi_{\f}(\B(\Y_{\uni},r))$ denote the total number of points of $\Phi_{\f}$ falling in ball $\B(\Y_{\uni},r)$. Therefore, $\mathbb{P}(\Phi(\B(\Y_\uni,r))=0)$ denotes the void probability \ie~ the probability that no point of $\Phi_{\f}$  falls in the ball $\B(\Y_{\uni},r)$. Therefore 
\begin{align*}
F_{R_{\Ct}|\Y_{\uni}}(r)&=1-\mathbb{P}\left(\Phi_{\f}(\B(\Y_\uni,r)=0|\Y_{\uni})\right)\nonumber
\end{align*}
	
\begin{theorem}
	\label{thm:1}
The CDF of the contact distance $R_{\Ct}$ of an event for a FHPPP with density $\lambda_\f$ is given as
\begin{align}\label{contactdistance}
F_{R_{\Ct}}(r)&=\1(r\leq2r_\drm)\left(1-\int_{y=0}^{r_\drm}e^{
	-\lambda_{\f} \min(\pi r^2,\pi r_\drm^2 ,\A {r}{r_\drm}{y})
}f_{\dist{\Y_{\uni}}}(y)\dv y\right)+\1(r\geq2r_\drm)(1-e^{-m}).
\end{align}
\end{theorem}
\begin{IEEEproof}
	See Appendix \ref{pthm1}.
\end{IEEEproof}
\begin{remark}
The case $r\geq 2r_\drm$ in Theorem \ref{thm:1} refers to the case where there is no sensor in the forest, and hence the distance to the nearest sensor is taken as $\infty$. This is an artifact of PPP that there is zero point of PPP in a finite range with certain probability.
\end{remark}

\subsection{CDF of the nearest neighbor sensor distance from a typical sensor}\label{thm2}

 The nearest neighbor sensor distance $R_{\Nt}$ is defined as the distance of a typical sensor to its nearest neighbor sensor. In the case of finite homogeneous PPP, the CDF of contact distance and nearest neighbor distance will be the same (See Appendix \ref{pthm2} for the proof). Therefore,
\begin{align}
F_{R_\Ct}(r)=F_{R_\Nt}(r),\,\, \forall r.\label{eq:nearestneighbor}
\end{align} 
Similar to the previous case, $r>2r_\drm$ refers to the scenario where there is no other sensor in the forest.

\begin{theorem}
	\label{thm3}
		The expressions for the upper bound $\overline{F_{R_\Ct}}(r)$, and the lower bound $\underline{F_{R_\Ct}}(r)$, on the CDF of the event contact distance  is given by:
	\begin{align}
	\overline{F_{R_\Ct}}(r)=&\1(r\leq 2r_\drm)\left(1-\left[A(r)+\frac{2}{\st r_{\drm}^2}\left(e^{-\st r}\left(r_{\drm}-\frac{1}{\st}\right)+e^{-\frac{\alpha^2(r)}{\lambda_{\f}}}
	\left(\frac{1}{\st}-|r_{\drm}-r|\right)\right)\right]\right),\\
	\underline{F_{R_\Ct}}(r)=&\1(r\leq2r_\drm)\left(1-\left[A(r)+\frac{2(r_{\drm}+r)}{r_{\drm}^2\sqrt{\lambda_{\f}}}\left(\erf\left(-\frac{r}{2} \sqrt{\lambda_{\f} \pi}\right) 
	-\erf
	\left(-\frac{\st}{2}  \sqrt{\frac{\pi}{\lambda_{\f}}}
	\right)
	\right)\right.\right.\nonumber\\
	&\left.\left.+\frac{4}{\pi r_{\drm}^2\lambda_{\f} }\left(e^{-\frac{\pi \alpha^{2}(r)}{4 \lambda_{\f}}}-e^{-\frac{\lambda_{\f} \pi r^2}{4}}\right)\right]\right)\label{lowerbound},
	\end{align}
where, 
\[\alpha(r)=2\lambda_{\f}\min(r,r_\drm),\   \text{ \ and} \]
\[A(r)=e^{-\lambda_{\f} \min(\pi r^2,\pi r_{\drm}^2)}\frac{(r_{\drm}-r)^2}{r_{\drm}^2}.\]
\end{theorem}
\begin{IEEEproof}
	See Appendix \ref{pthm3}.
\end{IEEEproof}

\begin{remark}
Another (loose) upper bound on the CDF of the event contact distance is given as:
\begin{align}\label{re:contactdistance}
\overline{F_{R_{\Ct}}}\le\overline{\overline{F_{R_{\Ct}}}}(r)&=\1(r\leq2r_\drm)\left(1-e^{
	-\lambda_{\f} \min(\pi r^2,\pi r_\drm^2 )
}\right)+\1(r>2r_\drm)(1-e^{-m}).
\end{align}
\end{remark}
\textit{Proof:} The upper bound $\overline{\overline{F_{R_{\Ct}}}}(r)$ can be achieved by replacing the intersecting area with its corresponding upper bound $\min(\pi r^2,\pi r_\drm^2)$.
\subsection{Asymptotic behavior of $F_{R_{\Ct}}(r)$ with $r_\drm$ while keeping $m$ fixed:}
\subsubsection{As $r_\drm\rightarrow0$} In this case, the  term $\min(\pi r^2,\pi r_\drm^2,\mathcal{A}(r,r_\drm,y))=\pi r_\drm^2$, therefore, the contact distance distribution will be:
\begin{align*}
F_{R_{\Ct}}(r)&=\lim_{r_\drm\rightarrow 0}\1(r\leq2r_\drm)\left(1-\int_{y=0}^{r_\drm}e^{\left(-\lambda_{\f} \pi r_\drm^2 \right)} f_{\Y_{\uni}}(y)\dv y\right)+\1(r> 2r_\drm)(1-e^{-m}).\\
&=1-e^{-m}.
\end{align*}
\subsubsection{As $r_\drm\rightarrow\infty$} In this case, the term  $\min(\pi r^2,\pi r_\drm^2,\mathcal{A}(r,r_\drm,y))=\pi r^2$. Hence, the contact distance distribution will be:
\begin{align*}
F_{R_{\Ct}}(r)&=\lim_{r_\drm\rightarrow \infty}\1(r\leq2r_\drm)\left(1-\int_{y=0}^{r_\drm}e^{
	-\lambda_{\f} \pi r^2 
} f_{\Y_{\uni}}(y)\dv y\right).\\
&=\lim_{r_\drm\rightarrow \infty}\1(r\leq2r_\drm)\left(1-\int_{y=0}^{r_\drm}e^{\left(-\lambda_{\f} \pi r^2 \right)} \frac{2y}{r_\drm^2}\dv y\right).\\
&=1-e^{-\lambda_{\f}\pi r^2}.
\end{align*}
When $r_\drm\rightarrow\infty$, FHPPP trivially converges to a homogeneous PPP with intensity $\lambda_{\f}$.
\section{Capacity functional of FHPPP and the event sensing probability}
Let the set $K_{\Y_\uni}(t)$ denote the envelop of an event that have occurred centered at location $\Y_{\uni}$. Recall the assumption that $\Y_{\uni}$ is uniformly located in $\B(\ob,r_\drm)$ and independent of FHPPP. We assume that the event envelop $K_{\Y_{\uni}}$ is expanding with time $t$.

 The event sensing probability $T_K(t)$ at time $t$ is the probability that the event envelop is sensed by at least one sensor of the WSN. Note that the event will be sensed if and only if the intersection of $K(t)$ with $\xi$ is non empty. Hence,
 \begin{align*}
 T_{K}(t)=\mathbb{P}(\xi \cap K_(t)\neq \phi).
 \end{align*}
 
 Similar to the previous section, we will start the derivation with conditioning on the location $\Y_\uni$. Conditioned on $\Y_\uni$, 
 $T_{K_{\Y_\uni}}(t)$ at time $t$, is the probability that an event started at center $\Y_\uni$ is sensed at time $t$.  Mathematically, it can be written as:
\begin{align*}
T_{K_{\Y_\uni}|\Y_{\uni}}(t)=\mathbb{P}(\xi \cap K_{\Y_\uni}(t)\neq \phi|\Y_{\uni}).
\end{align*}
Note that this is the capacity functional of finite Boolean-Poisson model evaluated at the set $K_{\Y_\uni}(t)$. 
\begin{theorem}\label{thm4}
The event sensing probability at time $t$ is given as
	\begin{align}
	T_{K}(t)=\int_{0}^{r_\drm}T_{K_{\Y_\uni}}(t)\frac{2y}{r_\drm^2}\dv y. \label{eq:finalcapacityfunctional}
	\end{align}
Here, $T_{K_{\Y_\uni}}(t)$ denotes the conditional event sensing probability at time $t$, conditioned on the starting location $\Y_{\uni}$ of  the event  and is given as
\begin{align}
	T_{K_{\Y_\uni}|\Y_{\uni}}(t)=1-\exp\left(-\lambda_{\f}\left|\B(\ob,r_\drm)\cap\left(\check{\St}\oplus K_{Y_\uni}(t)\right)\right|\right)
\end{align}
where  $\check{\St}$ is the complement set of $\St$ and $\St$ is $\B(\ob,r_\sen)$ and $\oplus$ denotes the Minkowski sum. Minkowski sum of any two set $\mathsf{A}\oplus\mathsf{B}$ is defined as $\{a+b: a\in\mathsf{A},b\in\mathsf{B}\}$.
\end{theorem}
\begin{IEEEproof}
	See Appendix \ref{pthm4}.
\end{IEEEproof}

\begin{remark}
	It can be easily seen that if $\mathsf{A}$ is a 2-dimensional ball then complement of $\mathsf{A}$ will be $\mathsf{A}$ itself.
	Therefore, $\check{\St}$ will be  $\B(\ob,r_\sen)$.
\end{remark}

\begin{remark}
	For the case of wild-fire, the $K_{\Y_{\uni}}(t)$ denotes fire-envelop which may take some shape depending upon the presence or absence of wind and its direction. In the absence of wind, the fire envelop expands with a velocity of $v_\F(t)$ in all directions. At time $t=0$, the fire envelop will be a point located at the point $\Y_{\uni}$ and at time $t$ it will become a circle with radius $v_\F(t) t$.
	
	Now, $\check{\St}\oplus K_{Y_\uni}(t)$ is the Minkowski sum of two balls which is equal to a ball of aggregate radius
	\begin{align}
	\check{\St}\oplus K_{Y_\uni}(t)=\B(\Y_{\uni},v_\F(t) t+r_\sen).
	\end{align}
	The fire sensing probability of a fire started at a typical point at time $t$ is given as:
	\begin{align}
	T_{K}(t)=1-
	\int_{0}^{r_\drm}e^{
		-\lambda_{\f}\left|\B(\ob,r_\drm)\cap\B(\Y_{\uni},v_\F(t) t+r_\sen)\right|
	}
	\frac{2y}{r_\drm^2}\dv y. \label{finalcapacityfunctional}
	\end{align}
\end{remark}

\begin{corollary}
	The coverage probability of a random point $\Y_{\uni}$ can be trivially achieved by putting $t=0$ and is given as
	\begin{align}
	\mathbb{P}[\xi\cap\{Y_{\uni}\}\neq& \phi]=1-\mathbb{P}[\xi\cap\{\Y_\uni \}=\phi]\nonumber \\
	T_{\Y_{\uni}}=&1-\int_{0}^{r_\drm}\exp\left(-\lambda_{\f}|\B(\ob,r_\drm) \cap \B(\Y_\uni,r_\sen)|\right)\frac{2 y}{r_\drm^2}\dv y\nonumber\\
	=&1-\int_{0}^{r_\drm}\exp\left(-\lambda_{\f}\mathcal{A}(r_\drm,r_\sen,y)\right)\frac{2 y}{r_\drm^2}\dv y.
	\end{align}
\end{corollary}
\begin{theorem}
	Bounds on the event sensing probability is given by:
	\begin{align}
	\overline{T_{K}}(t)&=\1(r_{\F}(t)\leq 2r_\drm)\left(1-\left[A\left(r_{\F}(t)\right)+\frac{2}{\alpha(t) r_{\drm}^2}	\left(e^{-\alpha(t) r_{\F}(t)}\left(r_{\drm}-\frac{1}{\alpha(t)}\right)+e^{-\frac{\alpha^2\left(r_{\F}(t)\right)}{\lambda_{\f}}}\left(\frac{1}{\alpha(t)}\right.\right.\right.\right.\nonumber\\
	&\vphantom{\int_1^2}\left.\left.\left.\left.-|r_{\drm}-r_{\F}(t)|\vphantom{\int_1^2}\right)\right)\right]\right),\\
	\underline{T_{K}}(t)&=\1(r_{\F}(t)\leq 2r_\drm)\left(1-\left[A(r_{\F}(t))+\frac{2(r_{\drm}+r_{\F}(t))}{r_{\drm}^2\sqrt{\lambda_{\f}}} \left(\erf\left(-\frac{r_{\F}(t)}{2} \sqrt{\lambda_{\f} \pi}\right)\right.\right.\right.\nonumber\\
	&\left.\left.\left.-\erf\left(-\frac{\alpha(t)}{2}  \sqrt{\frac{\pi}{\lambda_{\f}}}\right)\right)+\frac{4}{\pi r_{\drm}^2\lambda_{\f} }\left(e^{-\frac{\pi \alpha^{2}(t)}{4 \lambda_{\f}}}-e^{-\frac{\lambda_{\f} \pi r_{\F}^2(t)}{4}}\right)\right]\right)\label{lowerbound},
	\end{align}
\end{theorem}
where $\alpha(t)=2\lambda_{\f}\min(r_\F(t),r_\drm)$, $r_{\F}(t)=v_{\F}(t)t+r_{\sen}$ and $A(r_\F(t))=e^{-\lambda_{\f} \min(\pi r_{\F}(t)^2,\pi r_{\drm}^2)}\frac{(r_{\drm}-r_{\F}(t))^2}{r_{\drm}^2}$.

Another bound over the event sensing probability  is given by:
	\begin{align}
	&\overline{T_{K}}(t)\leq\overline{\overline{T_{K}}}(t)=1-\exp(-\lambda_{\f}\pi r_{\F}^2(t)).
	\end{align}

\subsection{Asymptotic analysis}
\subsubsection{As $r_\drm\rightarrow0$} In this case, the term $|\B(\ob,r_\drm)\cap(\check{\St}\oplus K_{Y_\uni}(t))|=\pi r_\drm^2$. Hence,
\begin{align*}
\lim_{r_\drm\rightarrow 0} T_{K}(t)&=1-e^{-m}
\end{align*}
\subsubsection{As $r_\drm\rightarrow \infty$} In this case, the term $\left|\B(\ob,r_\drm)\cap(\check{\St}\oplus K_{Y_\uni}(t))\right|=\left|\check{\St}\oplus K_{Y_\uni}(t)\right|$ which is not a function of $\Y_\uni$. Let us denote this term as $\left|\check{\St}\oplus K(t)\right|$.  Hence,
\begin{align}
T_{K}(t)=1-e^{-\lambda_{\f}|(\check{\St}\oplus K(t))|}
\end{align}
Therefore, the expression of the event sensing probability reduces to the capacity functional of Boolean-Poisson model as given in \cite{chiu2013stochastic}.

\section{Simulation results and analysis}
In this section, we present some numerical results to validate our analysis and provide insights about the system. 

\begin{figure}[ht!]
	\def\svgwidth{\textwidth}
	\centering
	\includegraphics[width=.46\textwidth]{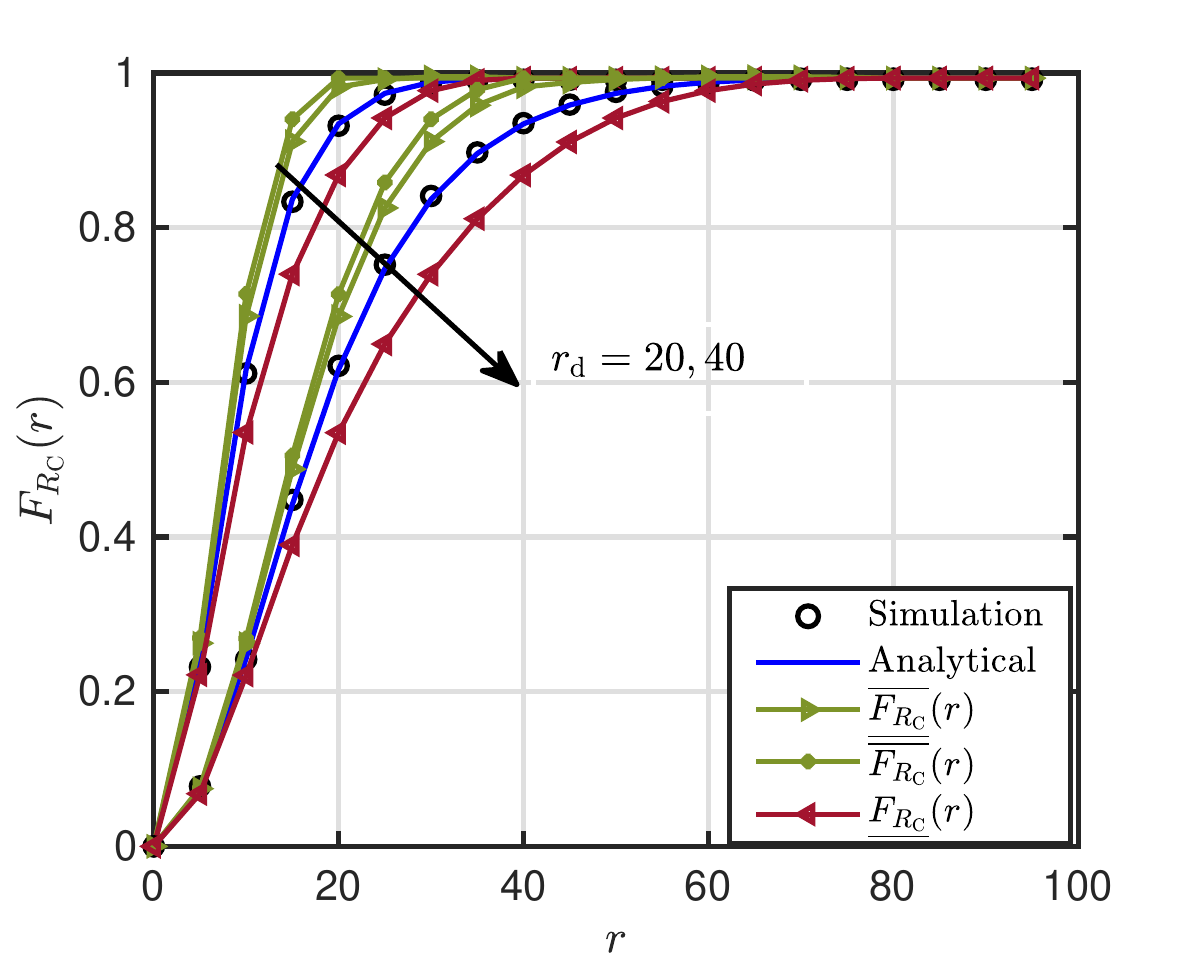}
	\caption{ Graph showing the CDF of nearest neighbor distance (or contact distance) and corresponding bounds for $m=5$. Increasing the $r_\drm$ will reduce the nearest neighbor distance distribution as the points will spread to far locations.  } \label{NNDoffiniteppp}
\end{figure}
\begin{figure}[ht!]
	\def\svgwidth{\textwidth}
	\centering
	\includegraphics[width=.46\textwidth]{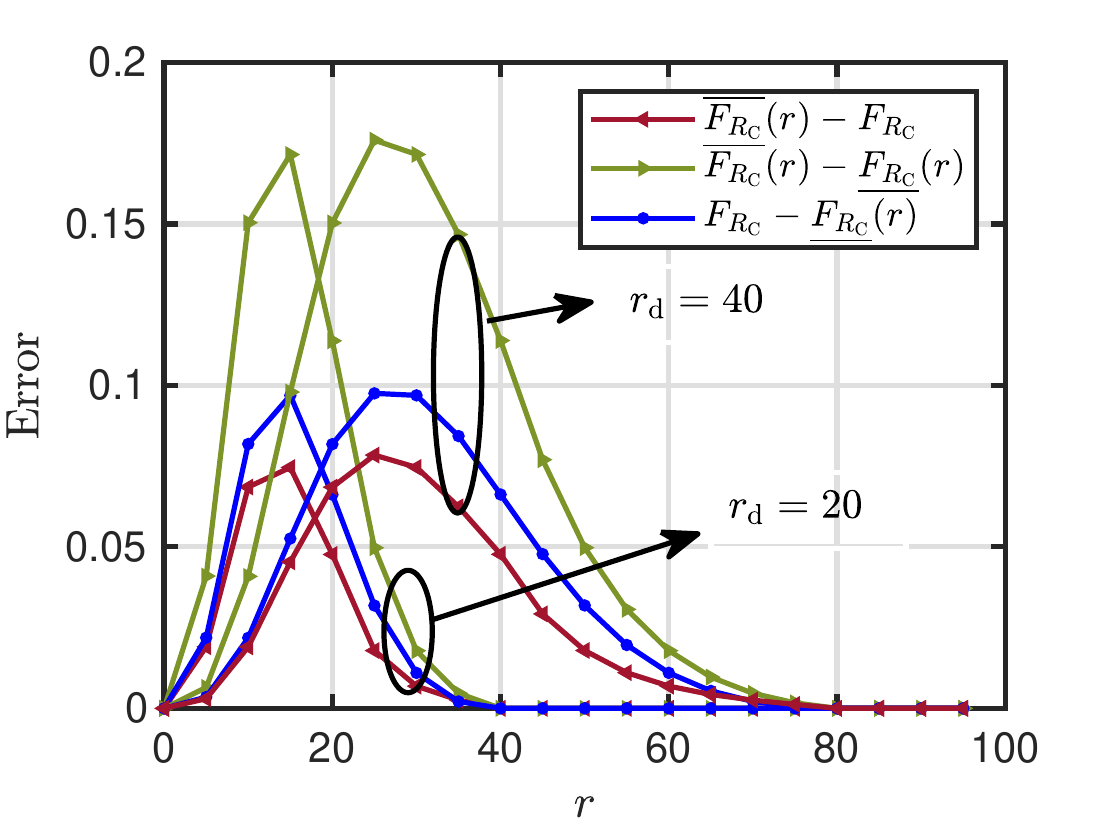}
	\caption{Deviation of the bounds from the exact values  for contact distance distributions. Here $m=5$.  } \label{errorcdfcontact}
\end{figure}

\subsubsection{CDF of contact distance and bounds}  

Fig. \ref{NNDoffiniteppp} shows the CDF of the nearest neighbor distance and corresponding upper and lower bounds for two values of $r_\drm$. It can be been easily seen that increasing $r_\drm$, will reduce the CDF because points will spread to far locations. Fig. \ref{errorcdfcontact} depicts the deviation of bounds from the exact values.

\subsubsection{Event sensing probability and corresponding bounds} 
\begin{figure}[ht!]
	\def\svgwidth{\textwidth}
	\centering
	\includegraphics[width=.46\textwidth]{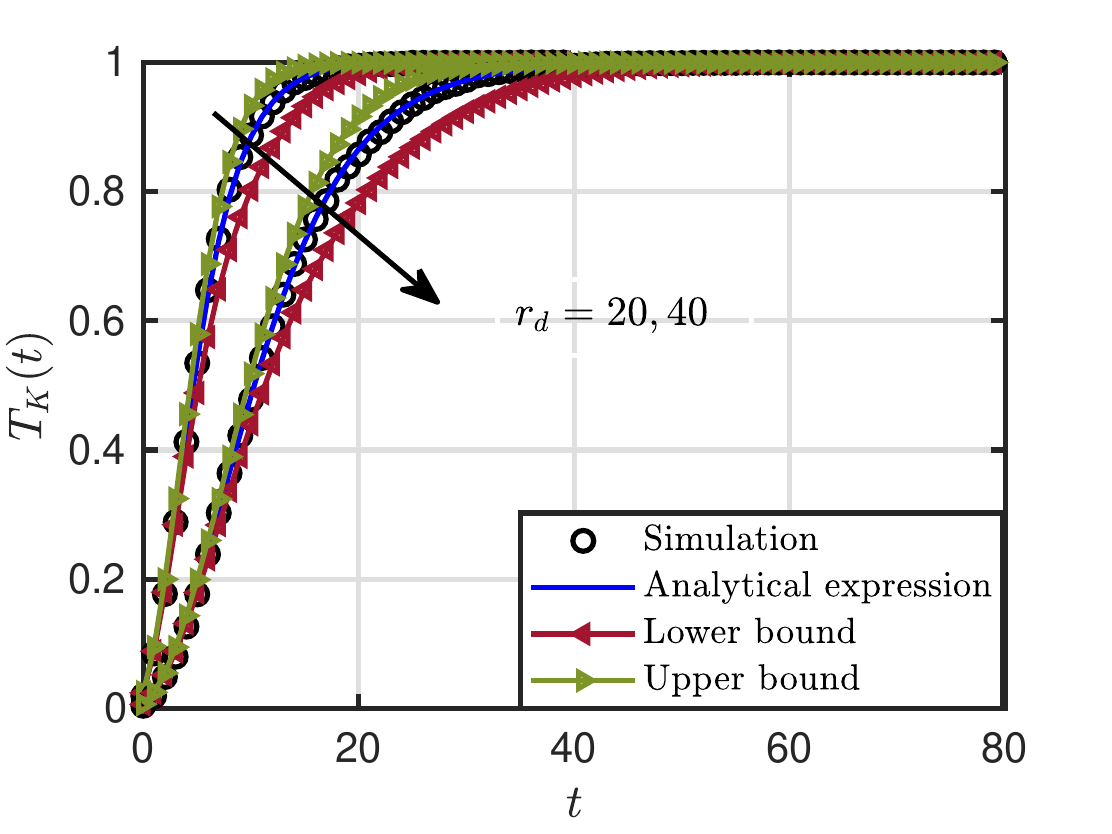}
	\caption{ The event sensing probability and corresponding deviation of bounds from exact values at time $t$ for wild-fires in the absence of  wind. The capacity functional (and hence sensing probability) increases with time $t$. Here, the  sensing range $r_{\sen}$ = 1 unit, $m=10$ and the flame velocity $v_{\F}=1 $ unit. } \label{fig:capacityfunctional1}
\end{figure}

\begin{figure}[ht!]
	\def\svgwidth{\textwidth}
	\centering
	\includegraphics[width=.46\textwidth]{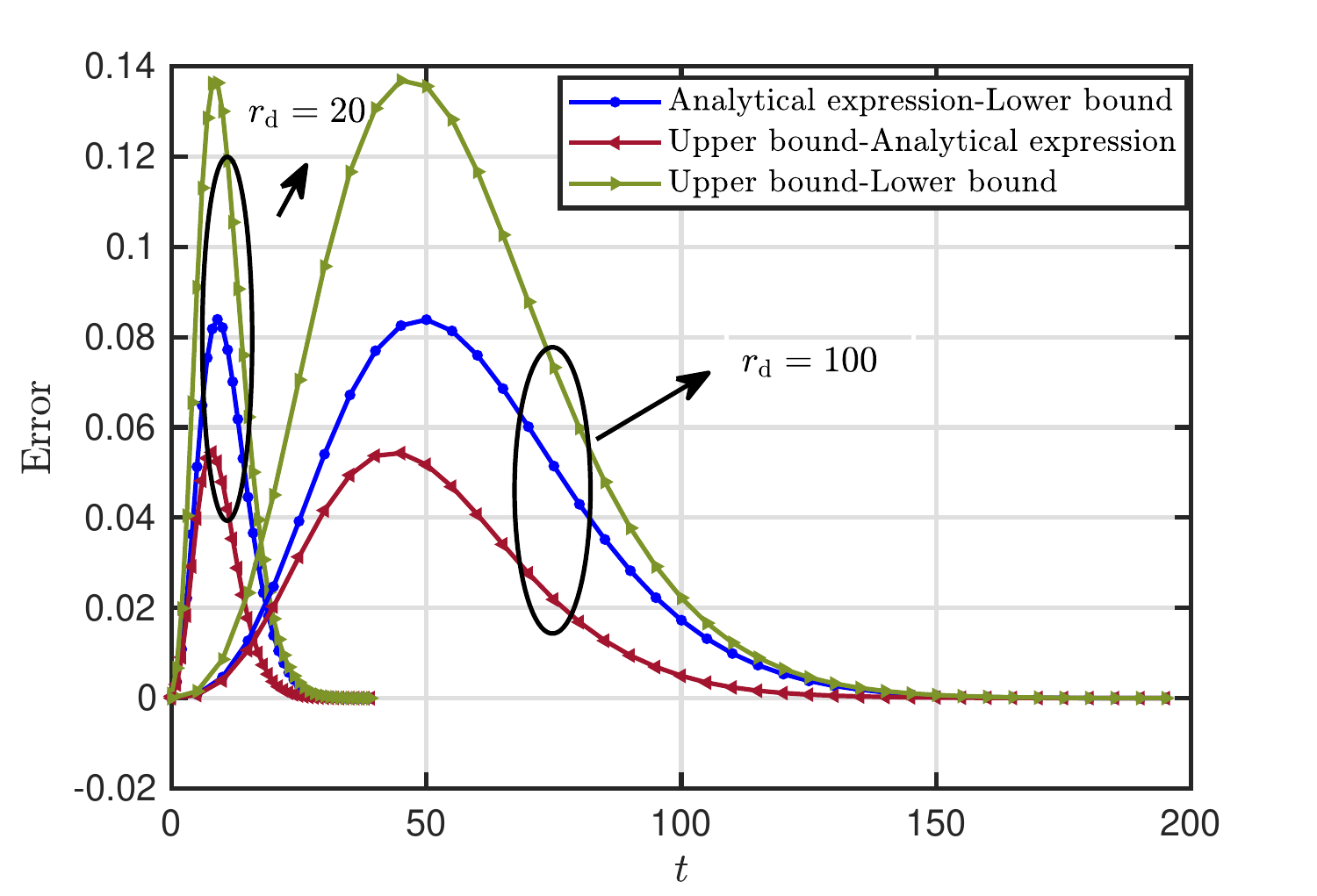}
	\caption{ The deviation of upper and lower bound for the sensing probability from exact values for two distinct values of the forest radius $r_\drm$.  Here, the  sensing range $r_{\sen}$ = 1 unit, $m=10$ and the flame velocity $v_{\F}=1 $ unit.  } \label{fig:capacityfunctional2}
\end{figure}

We now consider the case of wild-fires. Recall our assumption that fire envelop takes a circular shape with radius $v_\f(t)$.
Fig. \ref{fig:capacityfunctional1} shows the variation of fire sensing probability and the deviation of bounds from exact values  at time $t$. Intuitively, the fire sensing probability will increase with time. It is observed  from Fig. \ref{fig:capacityfunctional2} that the maximum error does not change much with respect to $r_\drm$. Therefore, the bounds may be tight even for higher values of $r_\drm$. 

\subsubsection{Impact of sensing range and number of sensors} 
\begin{figure}[ht!]
	\def\svgwidth{\textwidth}
	\centering
	\includegraphics[width=.46\textwidth]{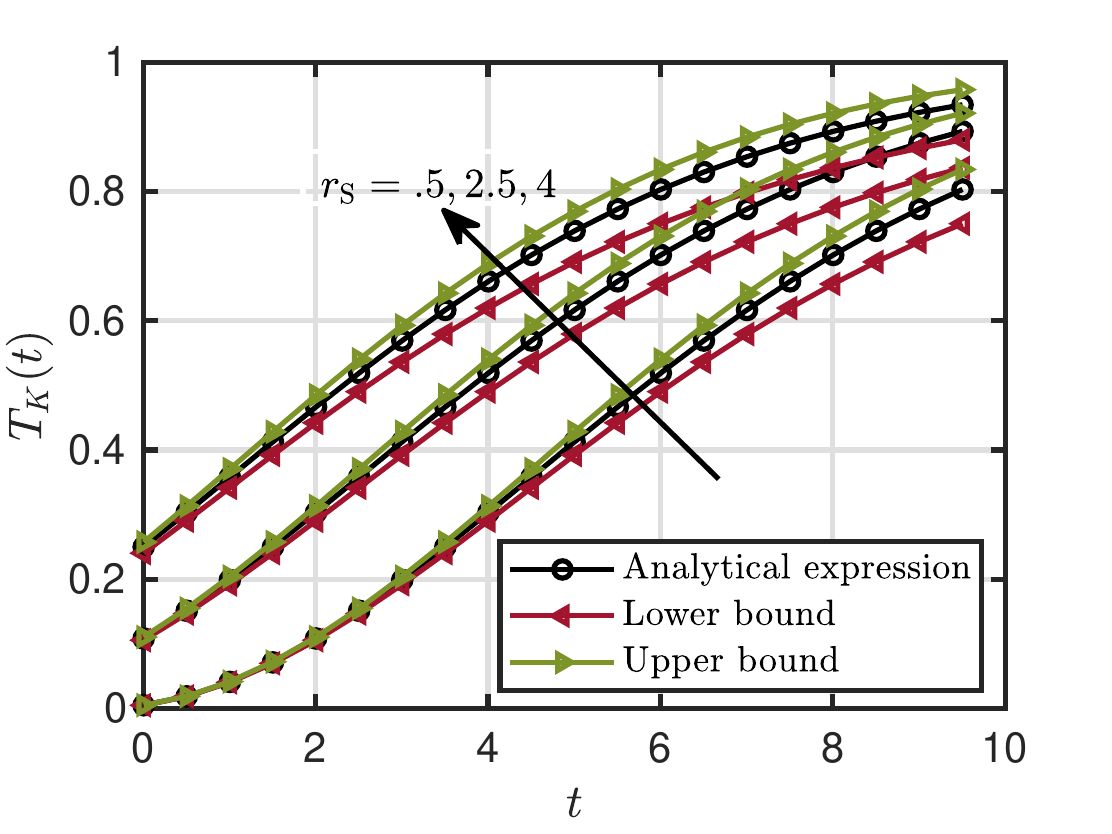}
	\caption{ Impact of sensing range on the fire sensing probability. Here, the forest radius of forest is 40 units. The mean number of sensors deployed are $40$. The fire flame velocity $v_\f$ is $0.5$ unit. Increasing the individual sensor's sensing can help in the early detection of wild-fires. } \label{cmptanalysis}
\end{figure}

\begin{figure}[ht!]
	\def\svgwidth{\textwidth}
	\centering
	\includegraphics[width=.46\textwidth]{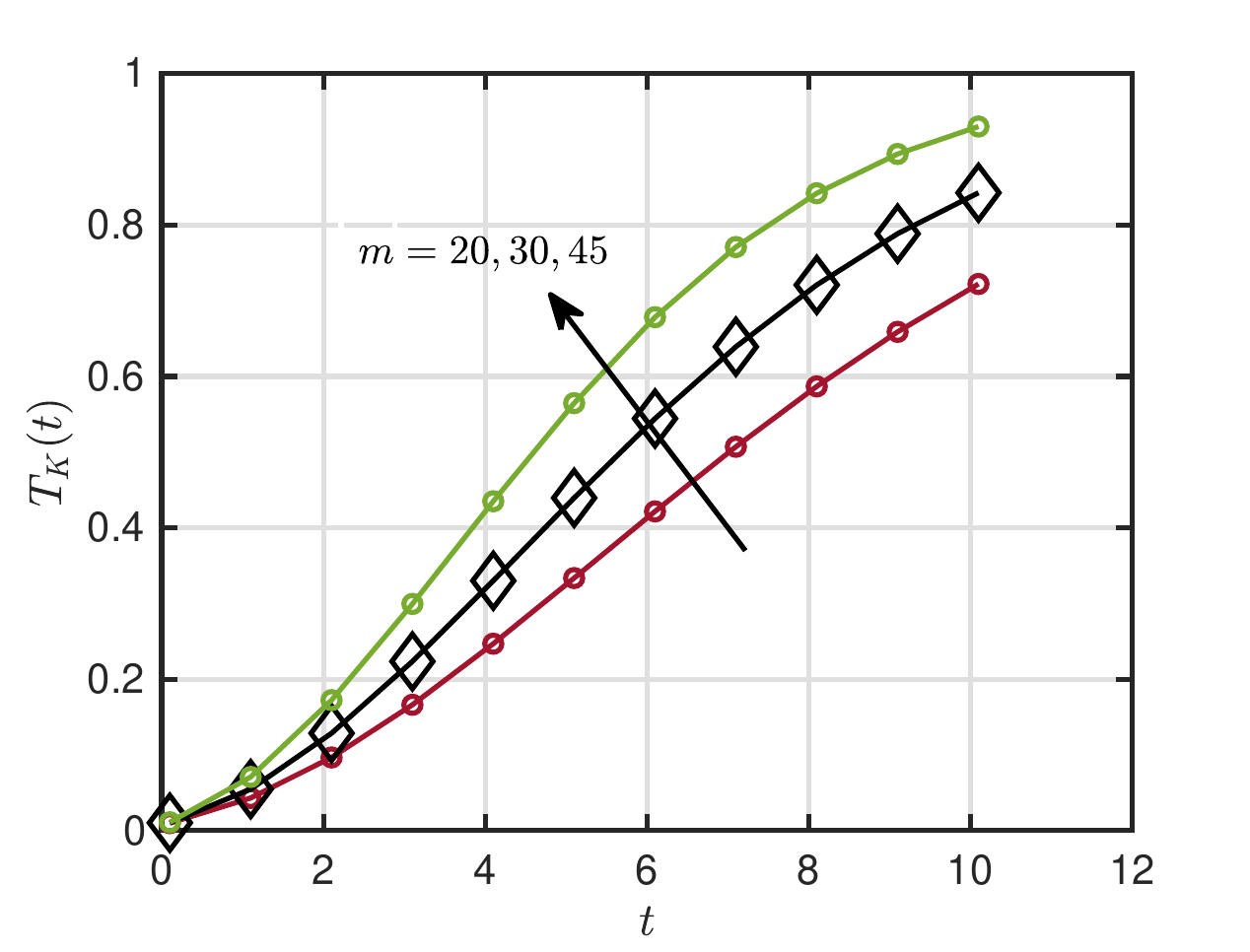}
	\caption{ Trade-off between the mean number of sensors ($m$) and the sensing range ($r_\sen$) for the fire sensing probability. We vary  $m$ and $r_\sen$ while keeping $m\pi r_\sen^2$  fixed at $40$. 
	The critical time before which the fire should be sensed is 10 unit. }\label{capacityfunctionalandsensingrangevarition}
\end{figure}

Fig. \ref{cmptanalysis} shows the impact of increasing sensing range on the fire sensing probability at time $t$ for a WSN with 40 sensors deployed in a forest with radius 40 units. We can observe that increasing the sensing range of sensors will increase the fire sensing probability. Hence, the critical time required to sense a fire with certain probability can be increased by increasing the sensing range which helps in early detection of fire. 
Fig. \ref{capacityfunctionalandsensingrangevarition} shows the trade-off between the mean number of sensors ($m$) and the sensing range ($r_\sen$) for the fire sensing probability, while keeping $m\pi r_\sen^2$ fixed. Note that $m\pi r_\sen^2$ denotes the sum of sensing areas of all sensors, and thus was used for the fair comparison. It can be observed that increasing the density of sensors have a higher impact on the fire sensing probability than increasing the individual sensor's sensing range. 
This can be justified in the following way. Increasing the number of sensors while reducing individual sensor's sensing range spreads the sensing region $\xi$ across the forest. On the other hand, increasing the individual sensor's sensing range while reducing the number of sensors localizes the sensing region $\xi$. 



\section{Conclusion}
This paper studies the dynamic event sensing performance of a randomly deployed wireless sensor network in a finite area (e.g. forest) modeled by a FHPPP. The paper analyzes the proximity of the wireless sensor network to an event, whose location  is uniformly distributed across the entire forest. In particular, we  study the event sensing probability  after time $t$  since the occurrence of the event. 
This paper presents an analytical expression for the CDF of contact distance and nearest neighbor distance distribution of FHPPP. The bounds presented are tight and can be used for the asymptotic analysis of the distance distribution and capacity functional. Finally, the simulation results validate the theoretical analysis. There are numerous possible extensions of this work. We have considered a circular fire propagation model which is an ideal case; there are several other fire propagation model exits in the literature. These fire propagation models are more realistic and can provide better insights into the sensor density. 
In some seasons the forest is prone to fire therefore the seasonal variations can also be included in the analysis to optimize the number of active sensors and to minimize the energy consumption of the sensor network.

\appendices
\section{Proof of Theorem \ref{thm:1}} \label{pthm1}
The  void probability of FHPPP for the set $\B(\Y_\uni,r)$ is given as
\begin{small}
	\begin{align*}
	&\mathbb{P}(\Phi_{\f}(\B(\Y_{\uni},r))
	=0|\Y_{\uni})=\mathbb{E}\left[\prod_{\X_i\in\Phi_{\f}}\1(\X_i\notin\B(\Y_\uni,r))\right]\\
	&\stackrel{(a)}=\exp\left(
		-\int_{\x\in
			\B\left(\ob,r_\drm\right)
			\cap\Phi_{\f}
		}
			\left(\vphantom{\frac12}1-\indside{\x\notin
				\B\left(\Y_\uni,r\right)}
			\right)
		\lambda(\x)\dv \x
		\right)\\
	&
	=\exp\left(-
	\int_{\x\in
		\left(\B(\ob,r_\drm)\cap\Phi_{\f}\right)}
	\indside{\X\in\B(\Y_{\uni},r)}
	\lambda(\x)\dv \x
	\right)\\
	&
	=\exp\left(-\lambda_{\f}
	\left|
	\B(\ob,r_\drm)\cap\B(\Y_{\uni},r)\right|
	\right)
	\end{align*}
\end{small}
Here, $(a)$ is due to the PGFL (probability generating functional) of finite PPP \cite{AndGupDhi16}. 
After  de-conditioning over $\Y_{\uni}$, we get  \eqref{contactdistance}.
\section{Proof of \eqref{eq:nearestneighbor}} \label{pthm2}
The distribution of the distance from the nearest neighbor sensor from any point is given as
\begin{small}
	\begin{align*}    
	F_{R_{\Nt}}(r)
	&=\frac{\mathbb{E}\left[\text{Number of sensors having nearest neighbor distance $\leq$ $r$}\right]}
	{\mathbb{E}\left[\text{Number of sensors}\right]
	}\\
	&=\frac{\mathbb{E}\left[\sum\limits_{\X_i\in \Phi_{\f}}\indside{R_{\Nt}(\X_i)\leq r}\right]}
	{\mathbb{E}\left[\sum\limits_{\X_i\in \Phi_{\f}}\1(\X_i \in \Phi_{\f})\right]}\\
	&\stackrel{(a)}=
	\frac1{\lambda_{\f}\pi r_\drm^2}
	{\int_{\x \in \B(\ob,r_\drm)}
		\lambda_{\f}\mathbb{P}^{\x}\left[R_{\Nt}(\x)\leq r\right]\dv \x}\\
	&=
	\frac1{\pi r_\drm^2}
	{\int_0^{r_\drm} \mathbb{P}^{\x!}\left[R_{\Ct}(\x)\leq r\right] 2 \pi x \dv x}\\
	&\stackrel{(b)}={\int_0^{r_\drm}
		\frac1{r_\drm^2}\ \mathbb{P}\left[R_{\Ct}(\x)\leq r\right]2 x\dv x}=\mathbb{E}_{\Y_\uni}\left[R_\Ct(\Y_\uni) \right]=F_{R_\Ct}(r)
	\end{align*}
\end{small}
Here, $(a)$ is due to Campbell Mecke's theorem. 
$\mathbb{P}^{\x!}$ is the reduced palm distribution and it is equal to the $\mathbb{P}$ for a PPP due to Slivnyak's theorem (step (b)). 

\section{Proof of Theorem \ref{thm3}} \label{pthm3}
In order to find the upper and lower bound of the contact distance $F_{R_\Ct}(r)$, consider the following integral:
\begin{align*}
f_1(r)=\int_{y=0}^{r_\drm}\exp{\left(-\lambda_{\f} \min(\pi r^2,\pi r_\drm^2 ,\A{r}{r_\drm}{y})\right)} f_{Y_{\uni}}(y)\dv y.
\end{align*}
For range $0\leq y\leq |r-r_\drm|$, the intersecting area $\A{r}{r_\drm}{y}=|\B(\ob,r_\drm) \cap\B(\Y_{\uni},r)|$ is equal to $\min(\pi r^2,\pi r_\drm^2)$ and the contribution of this range to the above integral is $e^{-\lambda_{\f}\min(\pi r^2,\pi r_\drm^2)}\frac{(r-r_\drm)^2}{r_\drm^2}$.

For range $|r-r_\drm|\leq y \leq r+r_\drm$, the intersecting area 
\begin{align}\label{areaofintersection}
\A{r}{r_\drm}{y}=&r_{\drm}^2\cos^{-1}\left(\frac{y^2+r_\drm^2-r^2 }{2yr_{\drm}}\right)
+r^2\cos^{-1}\left(\frac{y^2-r_\drm^2+r^2}{2yr}\right)\\&
-\frac12\sqrt{((r_\drm+r)^2-y^2)(y^2-(r_\drm-r)^2)}.
\end{align}
The integral for the range $|r_\drm-r|\leq y \leq r_\drm$  can not further simplified to its closed form. Hence, we will try to replace $\A{r}{r_\drm}{y}$ with its upper and lower bound.
\subsubsection{Proof of upper bound} 
Fig. \ref{boundarea}(A) shows the intersecting region  by the dotted area. The area of the rectangular shaded  serves as an upper bound for the area of the intersecting region. 
This rectangle has width $r+r_\drm-y$ and height $\min(2r,2r_\drm)$.  

\subsubsection{Proof of lower bound}
\begin{figure}[ht!]
	\def\svgwidth{\textwidth}
	\centering
	\includegraphics[width=.45\textwidth]{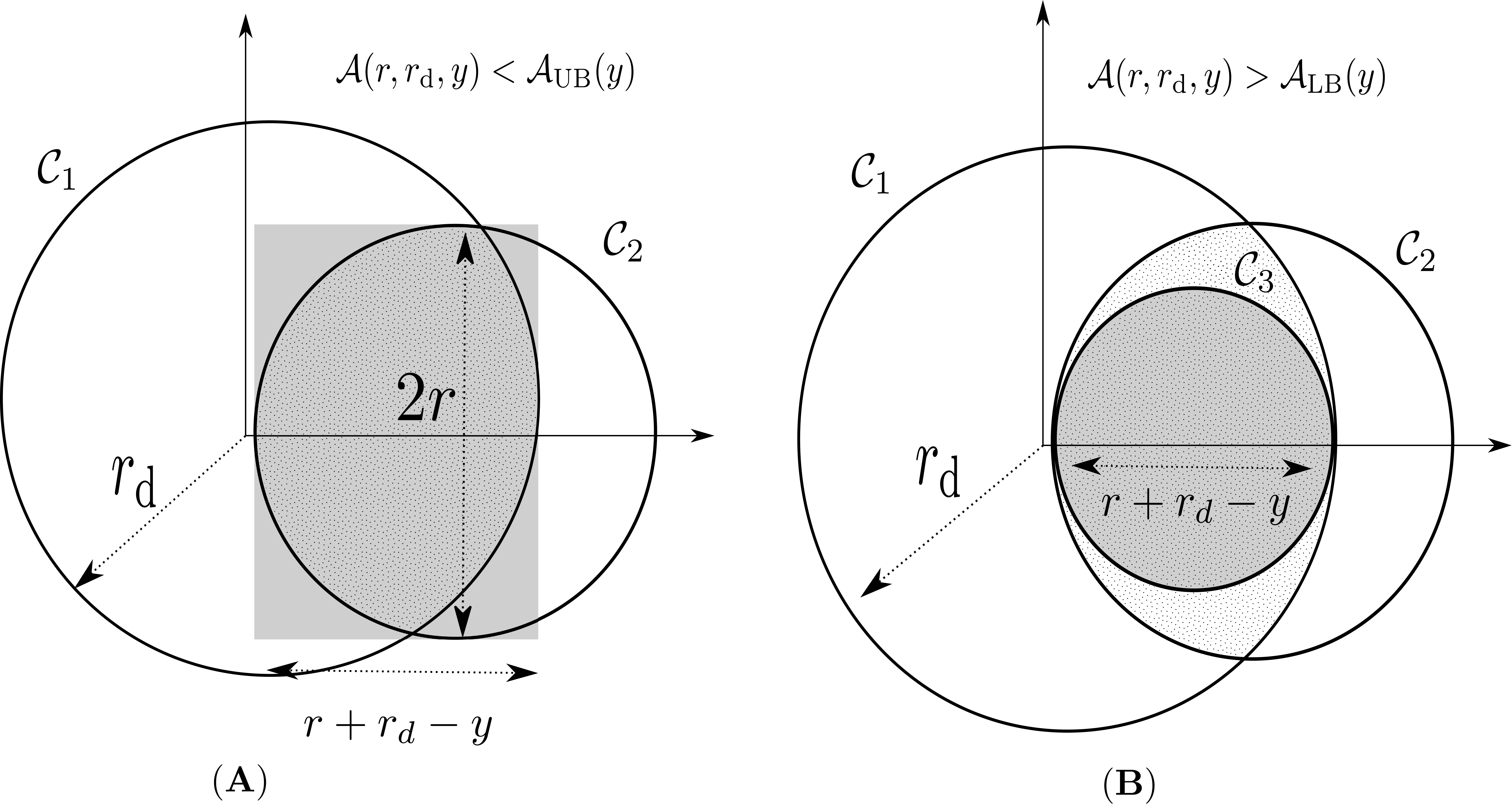}
	\caption{\small Illustration showing the upper and lower bounds for the intersecting area of two circles of radii $r_\drm$ and $r$  located at $y$ distance apart.} \label{boundarea}
\end{figure}

Let the two circle be $\C_1$  and $\C_2$, of radius $r_\drm$ and $r$ respectively.
Without loss of generality, assume $r_\drm>r$. 
Let the center of $\C_1$ is located at the origin. Let $\C_2$ have its center at $(\y,0)$. Therefore the distance between  center of the two circle is $y=||\y||$.  We have drawn a circle $\C_3$, of radius $\frac{r+r_{\drm}-y}{2}$ centered at $(\frac{r_\drm-r+y}{2},0)$.
It is clear from Fig.\ref{boundarea}(B) that circle $\C_3$ will touch $\C_1$ and $\C_2$ only at one point and the distance between the two is the diameter of $\C_3$. Therefore, $\C_3$ is completely under the intersecting region. Hence, its area serves as the lower bound for the intersecting area.
\section{Proof of Theorem \ref{thm4}} \label{pthm4} 
Conditioned on the occurrence point $\Y_\uni$, the event sensing probability is given as
\begin{align*}
&\mathbb{P}\left[\xi \cap K_{\Y_\uni}(t)\neq \phi|\Y_{\uni}\right]\\
&=1-\mathbb{P}\left[\xi \cap K_{\Y_\uni}(t)= \phi|\Y_{\uni}\right]\\
&=1-\mathbb{E}\left[\prod_{\X_i\in\B(\ob,r_\drm)\cap\Phi_{\f}}\1\left((\X_i+\St_i)\cap K_{\Y_\uni}(t)=\phi\right)\right]\\
&\stackrel{(a)}=1-\exp\left(-\lambda_{\f}\int_{ \B(\ob,r_\drm)}\left(\vphantom{\frac12}1-\1\left(\x \notin \check{\St}\oplus K_{\Y_\uni}(t)\right)\right)\dv \x\right)\\
&\stackrel{(b)}=1-\exp\left(-\lambda_{\f}|\B(\ob,r_\drm)\cap(\check{\St}\oplus K_{\Y_\uni}(t)) |\right).
\end{align*}
Here, $(a)$ is obtained from the  PFGL of FHPPP, and $(b)$ is due to the fact that integrating the product of indicators of two  sets over $\mathbb{R}^2$ results in the area of the intersection of these two sets.

\bibliographystyle{ieeetran}


\vspace{12pt}
\color{red}

\end{document}